\documentclass[seceq]{ptptex}

\usepackage{graphicx}

\title{
Chiral quark model approach for the study of baryon resonances %
}

\author{
Qiang \textsc{Zhao}%
}

\inst{
1) Institute of High Energy Physics, Chinese Academy of Sciences,
Beijing 100049, P.R. China \\
2) Theoretical Physics Center for Science Facilities,
  CAS, Beijing 100049, P.R. China}

\abst{
We report the chiral quark model approach for the study of baryon
resonances in meson photoproduction and meson-nucleon scattering
processes. Focussing on the $\pi N$ and $K N$ scatterings, we show
that the cross sections near threshold can be well accounted for by
the chiral quark model with only few parameters. The $S$-wave
dominance in both $\pi^- p\to \eta n$ and $K^- p\to \Sigma^0\pi^0$
can be well understood in the quark model framework. In particular,
the quark model provides a constraint on the relative signs for
resonance coupling form factors, which seems to be consistent with
results from isobaric models. }

\begin{document}

\maketitle

\section{Introduction}

The study of baryon resonances and search for so-called ``missing
resonances" in photo- and electroproduction, and meson-nucleon
scatterings have been one of the physics goals at international
hadron facilities. The increasing database gives access to more spin
observables by which detailed information about the underlying
dynamics can be extracted~\cite{Klempt:2005pp}. Theoretical analyses
are also making progresses on extracting baryon resonance parameters
(see e.g. proceeding of NSTAR2009 on which recent analyses from
EBAC~\cite{ebac}, SAID~\cite{SAID}, MAID~\cite{MAID},
Bonn-Gatchina~\cite{bonn-gatchina}, and Juelich~\cite{doering} are
reported). Meanwhile, it turns out to be urgent for theorists to
communicate with each other and understand the approaches employed
by different groups.

There is a clear need to treat all resonances consistently, and to
understand the relation between the {\it s}- and {\it u}-channel
resonances and {\it t}-channel meson exchanges. A recently developed
quark model framework~\cite{Li:1997gda}, augmented by an effective
Lagrangian approach to reaction dynamics, provides a good starting
point. The main feature of this model is the introduction of an
effective chiral Lagrangian for the quark-pseudoscalar-meson
coupling in a constituent quark model. In this framework,  the tree
level diagrams for pseudoscalar production reactions, such as in
meson photoproduction and meson-nucleon scattering, can be
explicitly calculated, and the quark model wavefunctions for the
nucleons and baryon resonances, after convolution integrals, provide
a form factor for the interaction vertices. Consequently, all the
{\it s}- and {\it u}-channel resonances can be consistently
included.

This model has the advantage of being able to describe a large
photoproduction database, employing only a very limited number of
parameters within a microscopic framework. Applications of this
model to the $\eta$~\cite{li-eta-95}, $K$~\cite{Li:1995sia}, and
pion photoproduction~\cite{li-pion,Zhao:2002id} have been quite
successful. Recently, this approach has been extended to
$\pi$-$N$~\cite{Zhong:2007fx} and $K$-$p$
scattering~\cite{Zhong:2008km} which provide some novel insights
into the observables measured in these two channels.

As follows, we give a brief introduction to the effective chiral
Lagrangian for quark-pseudoscalar-meson interactions in the SU(3)
flavor symmetry limit. We present results for $\pi^- p\to \eta n$
reaction in Sec.~\ref{pi-N}, and for $K^- p\to \Sigma^0\pi^0$ in
Sec.~\ref{K-N}. A summary is given in the last Section.

\section{The model}

In the chiral quark model, the low energy quark-meson interactions
are described by the effective Lagrangian
\cite{Manohar:1983md,Li:1997gda,Zhao:2002id}
\begin{eqnarray} \label{lg}
\mathcal{L}=\bar{\psi}[\gamma_{\mu}(i\partial^{\mu}+V^{\mu}+\gamma_5A^{\mu})-m]\psi
+\cdot\cdot\cdot,
\end{eqnarray}
where $V^{\mu}$ and $A^{\mu}$ correspond to vector and axial
currents, respectively.

The quark-meson pseudovector coupling at tree level can be extracted
from the leading order expansion of the Lagrangian
\begin{eqnarray}\label{coup}
H_m=\sum_j
\frac{1}{f_m}\bar{\psi}_j\gamma^{j}_{\mu}\gamma^{j}_{5}\psi_j\vec{\tau}\cdot\partial^{\mu}\vec{\phi}_m,
\end{eqnarray}
where $\psi_j$ represents the $j$-th quark field in the nucleon, and
$f_m$ is the octet pseudoscalar meson decay constant.

The $s$ and $u$ channel transition amplitudes are determined by
\begin{eqnarray}
\mathcal{M}_{s}=\sum_j\langle N_f |H_{m}^b |N_j\rangle\langle N_j
|\frac{1}{E_i+\omega_a-E_j}H_{m}^a|N_i\rangle,\\
\mathcal{M}_{u}=\sum_j\langle N_f |H_{m}^a
\frac{1}{E_i-\omega_b-E_j}|N_j\rangle\langle N_j | H_m^b
|N_i\rangle,
\end{eqnarray}
where $\omega_a$ and $\omega_b$ are the energy of the initial and
final state mesons, respectively. $H_m^a$ and $H_m^b$ represent the
quark-meson couplings at tree level described by Eq. (\ref{coup}).
$|N_i\rangle$, $|N_j\rangle$ and $|N_f\rangle$ stand for the
initial, intermediate and final state baryons, respectively, and
their corresponding energies are $E_i$, $E_j$ and $E_f$, which are
the eigenvalues of the NRCQM Hamiltonian.  The $s$ and $u$ channel
transition amplitudes have been worked out in the harmonic
oscillator basis in Refs. \citen{Zhong:2007fx,Zhong:2008km}.

There are a limited number of parameters introduced in this
approach. Apart from the SU(6)$\otimes$O(3) quark model parameters,
i.e.  $m_q = 330$ MeV and $\alpha=400 $ MeV, the overall
quark-pseudoscalar-meson coupling can be related to the hadronic
couplings by axial-vector current conservation. Then, the relative
strengths among $s$-channel baryon resonances will be determined by
the quark model symmetry. At tree level, we adopt the resonance
masses and widths from the Particle Data Group~\cite{pdg2008}. This
treatment violates the unitarity constraints on the amplitudes. In
principle, one should do coupled-channel calculations to restore the
unitarity of the amplitudes, where the resonance masses and widths
can also be calculated in the same framework. This is under
development at this moment.

For the $t$-channel, we considered the scalar meson exchange (e.g.
$a_0$ exchange in $\pi^- p\to \eta n$~\cite{Zhong:2007fx} and
$\kappa$ exchange in $K^- p\to \Sigma^0\pi^0$~\cite{Zhong:2008km}),
and vector meson exchange if it is allowed~\cite{Zhong:2008km}. This
will introduce additional parameters for the $t$-channel meson
couplings. But they can be constrained by independent processes.

\section{$\pi^- p\to \eta n$}\label{pi-N}

The $\pi^- p\to \eta n$ reaction should be an ideal channel for
testing the chiral quark model approaches apart from photoproduction
reactions. As it has been examined that the Goldberger-Treiman
relation~\cite{goldberger-treiman} is respected well in the chiral
quark model, this channel allows extraction of the $\eta NN$
coupling which can be compared with that determined in $\eta$
photoproduction. Furthermore, this channel is useful for
investigating the $S$-wave resonance interferences, i.e.
$S_{11}(1535)$ and $S_{11}(1650)$, by which the information about
the nature of these two states can be obtained.

\begin{figure}[htb]
\begin{center}
\begin{tabular}{ccc}
\includegraphics[scale=0.85]{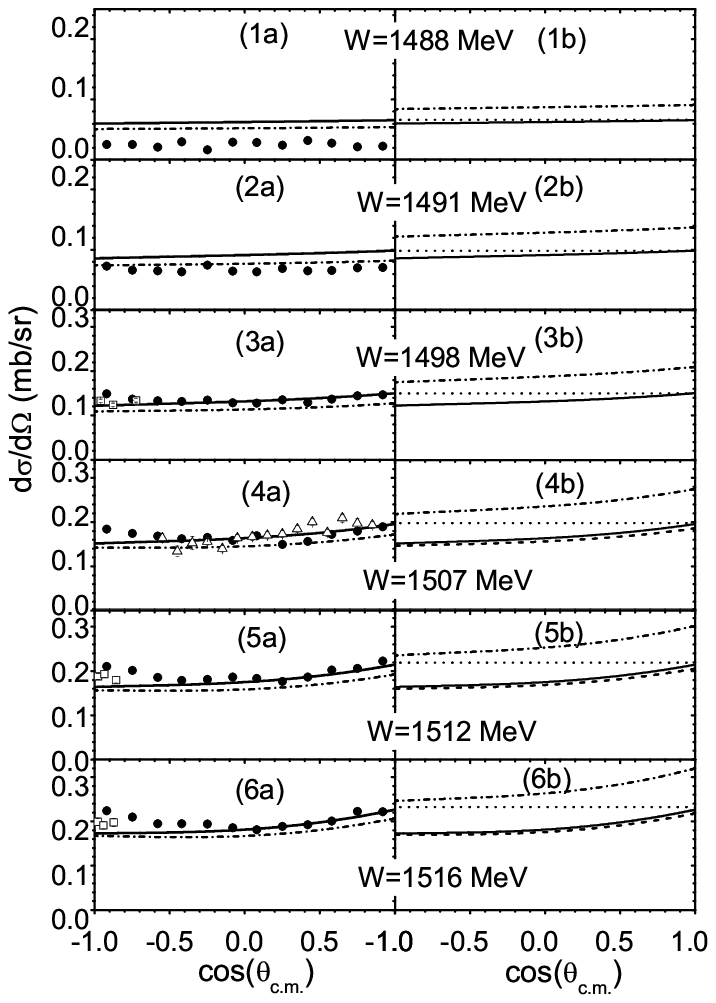}
\includegraphics[scale=0.85]{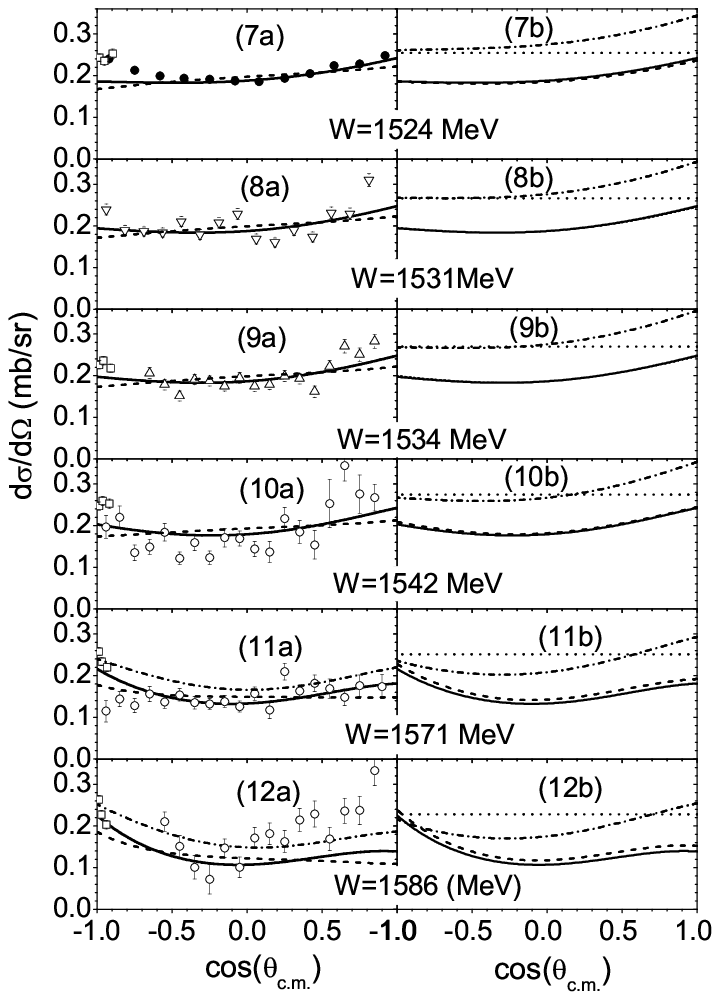}
\end{tabular}
\caption{The differential cross sections at various $W$. The data
are from \citen{exp1} (open circles), \citen{exp3} (open
up-triangles), \citen{exp4} (open down-triangles), \citen{exp6}
(open squares), and the recent experiment \citen{exp7} (solid
circles). The solid curves are for the full model differential cross
sections. In (1a-12a), the dash-dotted and dashed curves are for the
results switched off the contributions from nucleon pole and
$D_{13}(1520)$, respectively. In (1b-12b), the dash-dotted and
dashed curves correspond to the results without $S_{11}(1650)$ and
without $t$-channel, respectively; the straight lines corresponds to
the partial differential cross sections for $S_{11}(1535)$.
}\label{fig-1}
\end{center}
\end{figure}

In Fig. \ref{fig-1}, the differential cross sections near threshold
are illustrated and compared with experimental
data~\cite{exp1,exp3,exp4,exp6,exp7} (see the solid curves). We also
investigate effects arising from individual resonances as
illustrated in Fig. \ref{fig-1}. It shows that the interference
between $D_{13}(1520)$ and $S_{11}(1535)$ are crucial for producing
the correct shape for the differential cross section around the
$\eta N$ threshold. This feature is mentioned in
Refs.~\citen{1535,aa,Durand:2008es}, and similar feature also
appears in photoproduction
reactions~\cite{Tiator:1994et,li-eta-95,Chiang:2001as,Tiator:1999gr}.
The results also show that the $S$-wave contributions play a
dominant role near threshold. The enhanced cross sections after
removing the $S_{11}(1535)$ suggest that there exist cancellations
between the $S_{11}(1535)$ and other amplitudes. We shall see this
point clearer in the total cross sections.

\begin{center}
\begin{figure}[bt]
\includegraphics[scale=0.7]{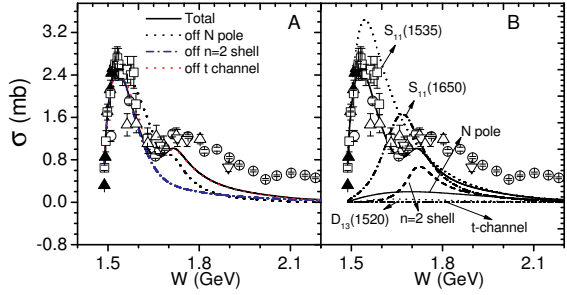} \caption{ The cross
section as a function of $W$. The data are from \citen{exp1} (open
circles), \citen{exp2} (open up-triangles), \citen{exp5} (open
down-triangles), \citen{Clajus:1992dh} (open squares), and the
recent experiment \citen{exp7} (solid triangles). The solid curves
correspond to the full model result, while the other lines represent
effects arising from individual resonances or single transition
amplitudes. }\label{fig-2}
\end{figure}
\end{center}

The total cross section turns out to be rather sensitive to the
interferences between those two $S$-wave states, i.e. $S_{11}(1535)$
and $S_{11}(1650)$. As shown by  Fig. \ref{fig-2}, in the region of
$W<1.6$ GeV (i.e., $p_\pi< 0.9$ GeV), the major contributions to the
cross sections are from the $S_{11}(1535)$ and $S_{11}(1650)$. The
contributions of the $S_{11}(1535)$ is about an order of magnitude
larger than those from the  $P$, $D$ and $F$ wave resonances. In
particular, it shows that the exclusive cross section from
$S_{11}(1535)$ is even larger than the data. But the destructive
interferences from the $S_{11}(1650)$ bring down the cross sections,
and as a consequence,  a ``second peak" around $W\sim 1.7$ GeV
appears in the total cross section.

\begin{table}[ht]
\begin{center}
\begin{tabular}{|c|c|c|c|c|c|}
  \hline
   $\langle\hat{H}\rangle$& $\Lambda K^+$
   & $p\eta$ & $ n\pi^+$ & $p\pi^0$ & $\Sigma^{+}K^0$ \\
   \hline
$\langle
N,J_z=\frac{1}{2}|\hat{H}|S_{11}^+(1535),J_z=\frac{1}{2}\rangle $&
$-\frac{1}{6}$& $-\frac{\cos{\theta}}{3\sqrt{3}}$ & $-\frac{2\sqrt{2}}{9\sqrt{3}}$ & $\frac{2}{9\sqrt{3}}$ & $-\frac{1}{9\sqrt{6}}$ \\
   \hline
$\langle
N,J_z=\frac{1}{2}|\hat{H}|S_{11}^+(1650),J_z=\frac{1}{2}\rangle $&
$0$ & $\frac{\cos{\theta}}{3\sqrt{3}}$ & $-\frac{\sqrt{2}}{9\sqrt{3}}$ & $\frac{1}{9\sqrt{3}}$& $\frac{2\sqrt{2}}{9\sqrt{3}}$ \\
   \hline
\end{tabular}
\caption{Spin-isospin factors for transition $S_{11}\to M N$. The
$S_{11}(1535)$ is assigned to SU(6)$\otimes$O(3) representation
$[{\bf 70, \ ^28}, \ 1, \ 1, 1/2^-]$, and $S_{11}(1650)$ to $[{\bf
70, \ ^4 8}, \ 1, \ 1, 1/2^-]$. Other charge conjugate channels can
be obtained by multiplying a proper isospin Clebsch-Gorden (C-G)
coefficient.}\label{tab-1}
\end{center}
\end{table}

Such a destructive interference could be a natural consequence of
non-relativistic constituent quark model (NRCQM) symmetry. In
Ref.~\citen{liu-zhao}, we show that based on the chiral Lagrangian,
one can determine the relative signs among $N^*NM$ couplings, where
$M$ and $N$ stand for pseudoscalar meson and octet ground state
baryon, respectively. The general form for the $S_{11}NM$ coupling
form factor can be expressed as
\begin{eqnarray}\label{trans-s}
{\cal M}_{S_{11}\to NM} &=& \frac{\langle\hat{H}\rangle}{f_m} [C_1
\alpha(q) +C_2 (\gamma(q)-\sqrt{2}\beta(q))],
\end{eqnarray}
where $C_1$ and $C_2$ are kinematic factors, and $\alpha(q)$,
$\gamma(q)$ and $\beta(q)$ are functions of final-state meson
momentum $q$ and given by the spatial integrals. These are the
common factors for $S_{11}(1535)$ and $S_{11}(1650)$ (if they are
degeneratte) in the symmetric NRCQM limit. The difference of the
coupling form factor is given by the spin-isospin factor $\langle
H\rangle$, which is listed in Table~\ref{tab-1} for different
coupling channels. One notices that for $\pi N\to S_{11}\to \eta N$,
there is a sign different arises from the spin-isospin factor
between $S_{11}(1535)$ and $S_{11}(1650)$ excitation amplitudes.
Details about the state mixings of these two $S_{11}$ states are
discussed in Ref.~\citen{liu-zhao}.

For $W > 1.6$ GeV, the contributions of $n=2$ resonances appear in
the reaction.  They play important roles around $W=1.7$ GeV. Without
the contributions from $n=2$ shell, the ``second peak" disappears.
It is possible that the $P$-wave states in $n=2$ shell have a
dominant contribution~\cite{shk,Ceci:2006ra}. To know which
resonance in $n=2$ shell contributes to the ``second peak", we
should rely on partial wave analysis, and more elaborate
consideration of SU(6)$\otimes$O(3) symmetry breaking within
individual resonances is needed.

\section{$K^- p\to \Sigma^0 \pi^0$}\label{K-N}

The reaction $K^-p\rightarrow \Sigma^0\pi^0$ is of particular
interest in the study of baryon resonances and $\bar{K}N$
interaction since there are no isospin-1 baryons contributing here
and it gives us a rather clean channel to study the $\Lambda$
resonances,  such as $\Lambda(1405)S_{01}$, $\Lambda(1670) S_{01}$,
$\Lambda(1520)D_{03}$ and $\Lambda(1690)D_{03}$. On the other hand,
the $K^- p\to \Sigma^0\pi^0$ reaction serves another opportunity to
examine the chiral quark model approach for the meson-nucleon
scatterings. Note that in the $s$-channel process the allowed
transitions will be via the processes that the initial and final
state meson couple to different constituent quarks. Therefore, the
$s$-channel will be qualitatively suppressed. This feature is
reflected by the much predominant contributions from the $u$-channel
amplitudes~\cite{Zhong:2008km}.

\begin{center}
\begin{figure}[ht]
\includegraphics[scale=3.2]{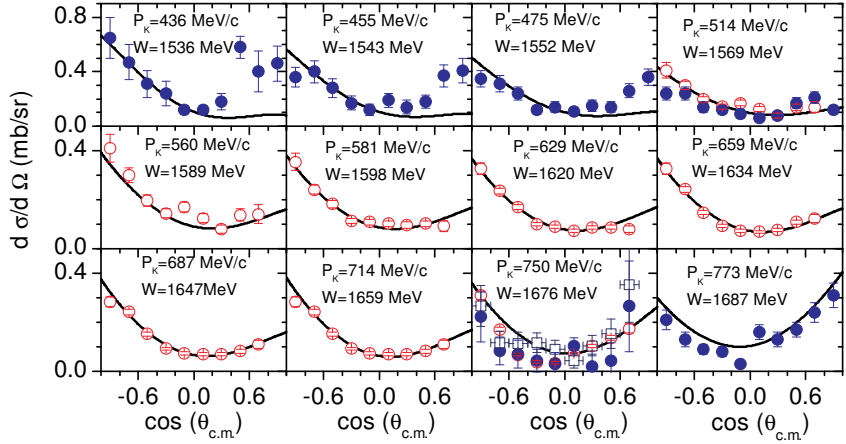} \caption{
Differential cross sections for $P_K=475\sim 775$ MeV/c (i.e.
$W=1536\sim 1687$ MeV). Data are from \citen{Manweiler:2008zz} (open
circles), \citen{Baxter:1974zs} (open squares) and
\citen{armen:1970zh} (round dottes). }\label{fig-3}
\end{figure}
\end{center}

In Fig. \ref{fig-3}, the differential cross sections are shown at
different center mass energies (beam momenta) from $W=1536$ MeV
($P_K=436$ MeV/c) to $W=1687$ MeV ($P_K=773$ MeV/c) in comparison
with experimental
data~\cite{Manweiler:2008zz,armen:1970zh,London:1975av,Baxter:1974zs,Berley:1996zh,Mast:1974sx}.
The solid curves denote full calculations and appear to have an
overall agreement with the data.

There are no data  for the differential cross sections available
near threshold, i.e. $W=1457\sim 1532$ MeV (or $P_K=200\sim 425$
MeV/c). Our calculation suggests that it is the region dominated by
the low-lying $\Lambda(1405)S_{01}$. In Ref.~\citen{Zhong:2008km},
exclusive cross sections by single resonance excitations or
transitions are presented. We find that the exclusive cross section
from $\Lambda(1405)S_{01}$ even overshoots the data near threshold
in the symmetric quark model limit. This feature is similar to the
observation in $\pi^- p$ scattering where the near-threshold region
is also dominated by the $S$-wave resonance excitations. Such a
large contribution from $\Lambda(1405)S_{01}$ would require a
cancellation mechanism, and hence the  $\Lambda(1405)S_{01}$ and
$\Lambda(1670)S_{01}$ mixing is favored~\cite{Zhong:2008km}. We note
again the similarity with $\pi^- p\to \eta n$ reaction.

\section{Summary}

The chiral quark model approach for the study of baryon resonances
in meson photoproduction and meson-nucleon scattering turns out to
have advantages for disentangling the $s$-channel resonances and
providing an estimate of background contributions on an equal basis
as the resonance amplitudes. In particular, the quark model provides
a natural constraint on the relative signs for resonance coupling
form factors, which can be compared with isobaric approaches. For
the dominance of the $S$-wave resonances, we show that a
cancellation appears between the first orbital excitation
$S_{11}(1535)$ and $S_{11}(1650)$ in $\pi^- p\to \eta n$, and
$\Lambda(1405)S_{01}$ and $\Lambda(1670) S_{01}$ in $K^- p\to
\Sigma^0\pi^0$. Such an effect could be essential for our
understanding of the constituent effective degrees of freedom within
excited baryons.

\section*{Acknowledgements}
The reported results are based on works carried out in collaboration
with X.-H. Zhong, J. He, B. Saghai, and X.-H. Liu. Q.Z. acknowledges
the hospitality of Yukawa Institute for Theoretical Physics. This
work is supported, in part, by the National Natural Science
Foundation of China (Grants No. 10675131), Chinese Academy of
Sciences (KJCX3-SYW-N2), and Ministry of Science and Technology of
China (2009CB825200).

%


\begin{thebibliography}{99}


\bibitem{Klempt:2005pp}
  E.~Klempt, C.~Batty and J.~M.~Richard,
  Phys.\ Rept.\  {\bf 413} (2005) 197.

\bibitem{ebac} H. Kamano, Chinese Phys. C 33 (2009), 1077.

\bibitem{SAID} R.A. Arndt, W.J. Briscoe, M.W. Paris, I.I.
Strakovsky, and R.L. Workman, Chinese Phys. C 33 (2009), 1063.

\bibitem{MAID} L. Tiator, D. Drechsel, S.S. Kamalov, M.
Vanderhaeghen, Chinese Phys. C 33 (2009), 1069.

\bibitem{bonn-gatchina} A. Sarantsev, Chinese Phys. C 33 (2009),
1085.

\bibitem{doering} M. D\"oring, C. Hanhart, F. Huang, S. Krewald, and
U.-G. Meissner, Chinese Phys. C 33 (2009), 1273.



\bibitem{Li:1997gda}
  Z.~P.~Li, H.~X.~Ye and M.~H.~Lu,
  Phys.\ Rev.\  C {\bf 56} (1997), 1099.

%
\bibitem{li-eta-95} Z.-P. Li,
    Phys. Rev. C {\bf 52} (1995), 4961;
    Z.-P.~Li and B.~Saghai,
    Nucl.\ Phys.\ A {\bf 644} (1998), 345;
  Q.~Zhao, B.~Saghai and Z.~P.~Li,
  J.\ Phys.\ G {\bf 28} (2002), 1293;
  B.~Saghai and Z.-P.~Li,
    Eur.\ Phys.\ J.\ A {\bf 11} (2001), 217.
%
\bibitem{Li:1995sia} Z.-P.~Li,
    Phys.\ Rev.\ C {\bf 52} (1995), 1648;
 Z.-P.~Li, W.~H.~Ma and L.~Zhang,
    Phys.\ Rev.\ C {\bf 54} (1996), 2171.
%
\bibitem{li-pion} Z.-P. Li,
    Phys. Rev. D {\bf 50} (1994), 5639.

\bibitem{Zhao:2002id}
  Q.~Zhao, J.~S.~Al-Khalili, Z.~P.~Li and R.~L.~Workman,
  Phys.\ Rev.\  C {\bf 65} (2002), 065204.

\bibitem{Zhong:2007fx}
  X.~H.~Zhong, Q.~Zhao, J.~He and B.~Saghai,
  Phys.\ Rev.\  C {\bf 76} (2007), 065205.

\bibitem{Zhong:2008km}
  X.~H.~Zhong and Q.~Zhao,
  Phys.\ Rev.\  C {\bf 79} (2009), 045202.
%



\bibitem{Manohar:1983md}
  A.~Manohar and H.~Georgi,
  Nucl.\ Phys.\  B {\bf 234} (1984), 189.

\bibitem{pdg2008}
  C.~Amsler {\it et al.}  [Particle Data Group],
  Phys.\ Lett.\  B {\bf 667}, 1 (2008).


%
\bibitem{goldberger-treiman}
  M.~L.~Goldberger and S.~B.~Treiman,
  Phys.\ Rev.\  {\bf 110} (1958), 1178.



\bibitem{exp1} R. M. Brown et al., Nucl. Phys. B \textbf{153}(1979),  89.


\bibitem{exp3} W. Deinet {\it et al.}, Nucl. Phys. B {\bf 11} (1969), 495.

\bibitem{exp4} J. Feltesse et al., Nucl. Phys. B {\bf 93} (1975), 242.


\bibitem{exp6} N.~C.~Debenham {\it et al.}, Phys.\ Rev.\  D {\bf 12}
(1975), 2545.


\bibitem{exp7} S. Prakhov et al., Phys. Rev. C {\bf 72}
(2005), 015203.


\bibitem{aa} A. M. Gasparyan, J. Haidenbauer, C. Hanhart, and J.
Speth, Phys. Rev C {\bf 68} (2003), 045207.

\bibitem{Durand:2008es}
  J.~Durand, B.~Julia-Diaz, T.~S.~Lee, B.~Saghai and T.~Sato,
  Phys.\ Rev.\  C {\bf 78} (2008), 025204.


\bibitem{1535} R. A. Arndt {\it et al.},  Phys. Rev C {\bf 72} (2005), 045202.
%

\bibitem{Tiator:1994et}
  L.~Tiator, C.~Bennhold and S.~S.~Kamalov,
  Nucl.\ Phys.\  A {\bf 580} (1994), 455.

%
\bibitem{Chiang:2001as}
  W.~T.~Chiang, S.~N.~Yang, L.~Tiator and D.~Drechsel,
  Nucl.\ Phys.\  A {\bf 700} (2002), 429.
%

\bibitem{Tiator:1999gr}
  L.~Tiator, D.~Drechsel, G.~Knochlein and C.~Bennhold,
  Phys.\ Rev.\  C {\bf 60} (1999), 035210.


\bibitem{liu-zhao} X.-H. Liu and Q. Zhao, work in progress.

\bibitem{exp2} F. Bulos et al., Phys. Rev. {\bf 187} (1969), 1827.

\bibitem{exp5} W. B. Richards et al.: Phys. Rev. D {\bf 1} (1970),  10.


\bibitem{Clajus:1992dh}
  M.~Clajus and B.~M.~K.~Nefkens,
  PiN Newslett.\  {\bf 7} (1992), 76.




\bibitem{shk} V. Shklyar, H. Lenske, U. Mosel, nucl-th/0611036.


\bibitem{Ceci:2006ra}
  S.~Ceci, A.~Svarc and B.~Zauner,
  Phys.\ Rev.\ Lett.\  {\bf 97} (2006), 062002.

\bibitem{Manweiler:2008zz}
  R.~Manweiler {\it et al.},
  Phys.\ Rev.\  C {\bf 77} (2008), 015205.


\bibitem{armen:1970zh}
R.~Armenteros {\it et al.},
Nucl.\ Phys.\  B {\bf 21} (1970), 15.


\bibitem{London:1975av}
G.~W.~London {\it et al.},
Nucl.\ Phys.\  B {\bf 85} (1975), 289.

\bibitem{Baxter:1974zs}
  D.~F.~Baxter {\it et al.},
  Nucl.\ Phys.\  B {\bf 67} (1973), 125.


\bibitem{Berley:1996zh}
  D.~Berley {\it et al.},
  Phys.\ Rev.\  D {\bf 1} (1970), 1996
  [Erratum-ibid.\  D {\bf 3} (1971), 2297].


\bibitem{Mast:1974sx}
  T.~S.~Mast, M.~Alston-Garnjost, R.~O.~Bangerter, A.~S.~Barbaro-Galtieri, F.~T.~Solmitz and R.~D.~Tripp,
  Phys.\ Rev.\  D {\bf 11} (1975), 3078.





\end{thebibliography}
\end{document}